\documentclass[compsoc, conference, a4paper, 10pt, times]{IEEEtran}

\usepackage{cite}
\usepackage{amsmath,amssymb,amsfonts}
\usepackage{algorithmic}
\usepackage{graphicx}
\usepackage{textcomp}
\usepackage{xcolor}
\usepackage{booktabs}
\usepackage[hidelinks]{hyperref}
\usepackage{makecell}
\usepackage{multirow}
\usepackage{caption}
\usepackage{subcaption}
\usepackage{xurl}
\begin{document}

\title{Attacker Profiling Through Analysis of Attack Patterns in Geographically Distributed Honeypots}

\author{\IEEEauthorblockN{Veronica Valeros}
\IEEEauthorblockA{\textit{Department of Computer Science}  \\ \textit{Czech Technical University} \\
\textit{Prague, Czech Republic}\\
veronica.valeros@fel.cvut.cz}

\and
\IEEEauthorblockN{Maria Rigaki}
\IEEEauthorblockA{\textit{Department of Computer Science}  \\ \textit{Czech Technical University} \\
\textit{Prague, Czech Republic}\\
maria.rigaki@fel.cvut.cz \\
}
\and
\IEEEauthorblockN{Sebastian Garcia}
\IEEEauthorblockA{\textit{Department of Computer Science} \\ \textit{Czech Technical University} \\
\textit{Prague, Czech Republic}\\
sebastian.garcia@agents.fel.cvut.cz \\
}
}

\maketitle

\begin{abstract}
Honeypots are a well-known and widely used technology in the cybersecurity community, where it is assumed that placing honeypots in different geographical locations provides better visibility and increases effectiveness. However, how geolocation affects the usefulness of honeypots is not well-studied, especially for threat intelligence as early warning systems. This paper examines attack patterns in a large public dataset of geographically distributed honeypots by answering methodological questions and creating behavioural profiles of attackers. Results show that the location of honeypots helps identify attack patterns and build profiles for the attackers. We conclude that not all the intelligence collected from geographically distributed honeypots is equally valuable and that a good early warning system against \textit{resourceful attackers} may be built with only two distributed honeypots and a production server. 
\end{abstract}

\begin{IEEEkeywords}
honeypots, early warning systems, geolocation, cybersecurity
\end{IEEEkeywords}

\section{Introduction}
Despite the long-standing tradition of using honeypots in computer security, our community still does not fully understand the relationship between the geographic location of honeypots (targets) and attacks. Many papers describe the properties of attacks on a honeypot~\cite{alata2006collection}, but there is a lack of research on the attack patterns (or the lack of patterns) towards multiple geographically distributed honeypots. 

There is a common assumption that geographically distributed honeypots help detect more attacks and attackers because the location may influence visibility~\cite{safarik2013automatic, gobel2010towards}. However, attack patterns seem complex to understand and predict, and the exact value of the information gathered is unclear at best~\cite{fraunholz2017data}. Understanding these patterns may significantly help the design of threat intelligence collection and their alleged utility as early warning systems.

Honeypot technology has been used successfully for many purposes, and extensive research has been done analysing the IP addresses of attackers, which ports are attacked, and the type of attacks~\cite{zhan2013characterizing, song2011statistical}. These analyses show that the number of attacks has varied over time and that there are many different types. Even some attack propagation models have focused on how the attackers attack~\cite{bar2016identifying}. However, no deep analysis has been done to find and understand attack patterns' long-term, stable properties beyond their numbers.

This paper explores attack patterns in geographically distributed honeypots by (i) using a public dataset of distributed honeypots for long-term, (ii) designing methodological questions to answer, (iii) processing data to extract results that answer the questions, (iv) creating behavioural profiles of attackers, (v) analysing the implications of the results.

Among the main questions asked are: Where do attacks originate from? How much do attackers attack? What do attackers attack? How do attackers attack? Can the attackers' behaviour be profiled? Does the geolocation of a honeypot impact the results obtained? Answers to these questions give rise to others, such as: What is the value of honeypots as early warning systems?

Results are disaggregated by attackers, attacks, profiles, number of honeypots and services attacked. From all attackers (source IP addresses), Asia has the larger amount with 54.87\% (almost as expected due to Internet population), and North America has 15.01\% (almost double as expected). In services attacked, 89\% of attackers attack only one port (10.99\% more than one). The highest \textit{median} number of attacks, 54, is for attackers that attacked more than 5 ports and all 8 honeypots, but the highest average number of attacks is for attackers that attacked more than 5 ports and only in 1 honeypot (2,222.12). Regarding the profiles of attackers, the highest number of \textit{attackers} (73.26\%) are the \textit{casual focused} (1 port, 1 honeypot), but the highest number of \textit{attacks} come from the \textit{driven explorers} ($<1$ honeypot and $>1$ port). In general, results show that Asia produces attacks as expected by its Internet population, Africa produces much fewer attacks, Europe produces slightly more, and North America produces more than twice the expected attacks by its Internet population.

We conclude that geographically distributed honeypots can play an important role in identifying attackers' profiles and attack patterns. This is because multiple honeypots help correlate and separate the \textit{casual} attacker from the \textit{driven} one. Honeypots can be used as early warning systems but \textbf{not} all the intelligence collected is equally essential. We conclude that with only two distributed honeypots and a production system, organisations can defend against the most aggressive and driven attackers while ignoring the vast majority of casual attackers.

The main contributions of this paper are:
\begin{itemize}
    \item Creation of four \textit{attacker profiles} that identify \textit{how} they attack based on the number of services and honeypots contacted.
    \item Identification of two attack patterns by driven explorer attackers, those attacking in ascending order and those in descending order, allowing the further study of honeypots as early warning systems.
    \item Determination of the minimum number of geographically distributed honeypots required to identify the most resourceful attackers to block in production services.
    \item Identification of a specialisation trend in global Internet attackers, where attackers prefer a narrow approach to attack certain services instead of multiple services at once.
    \item Identification of which attackers' profiles are valuable when honeypots are used as early warning systems.
\end{itemize}

The rest of the paper is organised as follows: Section~\ref{sec-prevwork} discusses existing research in the area; Section~\ref{sec-dataset} presents the details of the dataset to be studied; Section~\ref{sec-methodology} explains the questions to be explored and tools and methods used to answer those questions; Section~\ref{sec-results} presents the results for questions and proposed profiles; Section~\ref{sec-analysis} presents our interpretation and analysis of the results; Section~\ref{sec-recommendations} proposes some open recommendations to cybersecurity defenders; Section~\ref{sec-futwork} discusses limitations and future work; Section~\ref{sec-conclusions} presents the conclusions of this research.

\section{Previous Work}
\label{sec-prevwork}


In the past few years, several studies have proposed the use of honeypots placed in multiple geographical regions, mostly using cloud-based deployments. Some studies were narrowly focused on emulating only one service, such as SSH~\cite{sophos2021} or TELNET~\cite{strohmeier2021studying}. In contrast, others focused on specific domains such as the Internet of Things (IoT)~\cite{srinivasa2022interaction, strohmeier2021studying} or Industrial Control Systems (ICS)~\cite{using2020dodson}.

Srinivasa et al.~\cite{srinivasa2022interaction} extended the RIoTPot honeypot and emulated six different protocols related to IoT. The deployment period was three months. The geographical dispersion of their honeypots was limited to four cities since the main focus of the study was the comparison between the effect of the different types of honeypot interaction (low, medium, and high). The study found that high-interaction honeypots receive more attacks over time, but the distribution of attack types (brute-force, port scans, etc.) was similar. The dataset is shared on request. 

Kelly et al.~\cite{kelly2021comparative} used ten different honeypots that emulated multiple services and an Intrusion Detection System (IDS). The honeypots were deployed in three cloud providers and three geographical regions for three weeks. However, most honeypots were placed on the east coast of the United States, and only two were outside the US (one in the United Kingdom and one in Singapore). One of the main questions the authors tried to address was whether the attackers prefer to attack services hosted in cloud providers that have a higher market share. Their analysis showed no distinction between providers when it comes to attacking frequency, attacker geolocation, and targeted vulnerabilities. The dataset was not published.

A similar study was conducted by Priya et al.~\cite{priya2023containerized}. The study was limited to ten days and to only two geographical regions (India and Japan). The authors focused mainly on analysing DDoS attacks, including any post-exploitation commands executed in the honeypots for downloading the malware responsible for the attacks. The dataset was published, but network data is sparse.

SSH and TELNET services are very popular attack targets due to the plethora of IoT devices using them and the number of vulnerabilities these devices have. Sophos, a cybersecurity company, published a study where they deployed SSH-only honeypots to ten different Amazon cloud locations for 30 days~\cite{sophos2021}. Their analysis focused on the usernames and passwords mostly frequently used by attackers and the IoT devices associated with the respective default credentials. A TELNET-based low interaction honeypot with weak credentials was the basis of the deployment in~\cite{strohmeier2021studying}. The authors deployed honeypots in 13 different countries, trying to answer questions regarding attack neutrality in cyberspace. Their main finding was that low-sophistication attackers do not consider the location of their targets and that their attacks are mostly automated. The dataset was not published. 

The only work that tries to identify attack propagation patterns between honeypots is from Bar et al.~\cite{bar2016identifying}. The authors propose to use Markov Chains and complex network analysis to model attack propagation. They deployed 267 honeypots worldwide using four different honeypot software emulating multiple services. The deployment lasted for a year. The work identified propagation patterns between "communities" of honeypots; however, these patterns are not actionable from a threat intelligence perspective. The dataset was not published.   

\section{Dataset}
\label{sec-dataset}

To analyse attackers and attacks across regions, we used the CTU Hornet 40 dataset~\cite{valeros2022hornet}. The dataset contains $4,747,081$ attacks, performed by $265,423$ unique attackers, to $8$ geographically-distributed honeypots, during $40$ days. The data was captured from April 23rd, 2021 to June 1st, 2021.~\autoref{tab:honeypots-overview} presents a summary of the dataset per geographical location, the number of unique source IP addresses, and the number of network flows per honeypot.

\begin{table}[tb]
\scriptsize
\renewcommand{\arraystretch}{1.5}
\caption{Summary of the CTU Hornet 40 dataset, for the complete 40 days, showing the total number of unique source IP addresses (attackers) that connected to each honeypot and the total number of flows received per honeypot (attacks).}
\begin{center}
\begin{tabular}{l c r r}
\toprule
\textbf{Honeypot Name}  & \textbf{City} & \makecell{\textbf{Unique}\\\textbf{Source IPs}} & \makecell{\textbf{Network}\\\textbf{Flows}} \\
\midrule
Honeypot-Geo-1 & Amsterdam       & 36,441 & 347,195   \\
Honeypot-Geo-2 & Bangalore       & 59,103 & 444,007   \\
Honeypot-Geo-3 & Frankfurt       & 83,254 & 1,399,437 \\
Honeypot-Geo-4 & London          & 60,273 & 1,169,506 \\
Honeypot-Geo-5 & New York        & 48,967 & 298,851   \\
Honeypot-Geo-6 & San Francisco   & 41,478 & 308,829   \\
Honeypot-Geo-7 & Singapore       & 71,891 & 352,572   \\
Honeypot-Geo-8 & Toronto         & 52,824 & 438,260   \\
\bottomrule
\end{tabular}
\label{tab:honeypots-overview}
\end{center}
\end{table}

The network flows are bidirectional, meaning they aggregate packets' features in both directions: from source IP to destination IP and vice versa. Eight Virtual Private Servers (VPS) were rented in the same VPS provider~\cite{digitalocean:online}. VPS servers were rented in eight cities: Amsterdam, Bangalore, Frankfurt, London, New York, San Francisco, Singapore, and Toronto. The cities were chosen to maximise the coverage of continents conditioned on the availability of the VPS provider. Each Linux-based VPS was configured to act as a passive honeypot without any service, receiving attacks without running any honeypot software. This was done to receive an unbiased sample per port attacked, not conditioned on the open ports~\cite{srinivasa2022interaction}.

The dataset was loaded into a free self-hosted instance of Splunk~\cite{tools:splunk}, a security information and event management system for analysis, data enrichment, generation of statistics, and graphs. The dataset was also processed using Pandas\cite{mckinney2010data} to extract attack paths, and it was visualised using the Gephi tool~\cite{bastian2009gephi} to analyse the attacks as directed connected graphs. The only post-processing done in the dataset was the removal of ARP packets since those only come from the VPS provider's local network.

\section{Methodology}
\label{sec-methodology}

Throughout this work, the following definitions were used for \textit{attack}, \textit{attacker} and \textit{service}:

\begin{itemize}
    \item \textbf{Attack}: an attack is defined as a network flow whose destination IP address matches one of the IPs of the honeypots.
    \item \textbf{Attacker}: an attacker is defined as the source IP address in an attack (flow).
    \item \textbf{Service}: a service is defined as the destination port of an attack (flow).
\end{itemize}

The rest of this section presents the main questions and methods used in this work.

\subsection{Where do attacks originate from?}
\label{sec-where-attacks-from}

One of the motivations behind the need to know where attacks come from may be related to the physical-world assumption that attacks coming from the IP addresses of a country would be related, or even supported, by the authorities, companies, or organisations of that country. It is also related to the need for attribution for attacks. From a technical perspective, a good attribution may be used to better understand and profile the alleged attackers. However, there are many situations in which the IP address of a country may not be related to the country itself. Examples include cloud providers that lease shared IP addresses, remotely controlled computers, and the creation of fake traces to be used as false flags.

To analyse the origin of attacks, we first propose the new concept $H$ for each attacker seen in the dataset. $H$ is defined as \textit{the number of unique honeypots attacked}. This new concept helps focus on the attackers that are interested in few honeypots or maybe do not have resources to attack more honeypots, from attackers that are seen in all honeypots sending a large number of flows.

In our case, $H$ is defined in the range $1<=H<=8$. Second, we looked up the continent where the IP address of the attacker was registered using the MaxMind Geo Lite~\cite{tools:maxmindgeolite} database. MaxMind uses seven continents and obtains the location by correlating various sources of data~\cite{tools:maxmindgeolite}. The precise location of an IP may be inaccurate, but at a continent level, it is more precise.

The number of \textbf{attackers per source continent} was calculated as the sum of attackers per source continent, disaggregated by $H$ (number of unique honeypots attacked). This measures how many \textit{attackers} originated from each continent. The number of \textbf{attacks per source continent} was calculated as the sum of attacks per source continent, disaggregated by the $H$. This measures how many \textit{attacks} originated from each continent.

The data received in the honeypots was compared to the expected traffic for the Internet population. For this comparison, we used the \textit{world percentages of Internet penetration rate per continent}~\cite{miniwatts_marketing_group_world_2023}. The values used for our comparison as of January 2023, are Africa=11.2\%, Asia=58\%, Europe=13.9\%, South America=9.9\%, North America=6.5\%, and Oceania=0.6\%. These values represent the \textit{expected} average amounts per continent if we assume that \textit{in average}, people from each continent attack approximately the same amount.

\subsubsection{Tor proxy attackers}
\label{sec-tor-exit-nodes}

In the case of attackers hiding their IP address behind proxy servers, VPNs, or Tor, the analysis is still valid since it focuses on \textit{where} attacks come from and not in which is the real country of the attacker. If a proxy network is used for many attacks, this is still valid information for the defender.

Tor is the most well-known and used proxy network with more than 2 million users~\cite{wecsr10measuring-tor}. It is well known that attackers often proxy their attacks through it~\cite{tor:malware-mevade,tor:malware-chewbacca}. To determine the number of attackers proxying their attacks through the Tor network, we obtained the list of IPs used as Tor exit nodes in the time frame of the dataset. Tor publishes monthly lists of exit nodes, each containing the list of IPs for that day. We downloaded the data for the months of April, May and June 2021. From these months, we kept only the data corresponding to the days of the dataset, from the 23rd of April to the 1st of June 2021. The resulting data were processed to extract the IPs for each day, which were then combined and deduplicated to form a unique list and compared to the dataset.


\subsection{How much do attackers attack?}
\label{sec-how-much-attackers-attack}

By measuring how much attackers attack, regarding the number of flows sent, it is possible to define policies that focus on the attackers that are more \textit{aggressive}, \textit{persistent} or \textit{resourceful}. This information helps prioritise resources and focus defence mechanisms.

To answer this question, we calculated the number of attacks and attackers, disaggregated by $H$ (number of unique honeypots attacked). The number of \textbf{attackers per $H$} was calculated as the sum of the number of attackers that attacked $H$ honeypots. The number of \textbf{attacks per $H$} was calculated as the sum of the number of flows per attacker that attacked $H$ honeypots. Additionally, we calculated the average, standard deviation and median of the number of attacks per attacker per $H$.

\subsection{What do attackers attack?}
\label{sec-what-attackers-attack}
There are two main motivations for studying which services attackers attack. First, it shows the focus and interest of attackers, maybe based on the availability of exploits. This alone sheds light on the attackers' trends. Second, if the defenders can choose the port numbers of their services, learning which ports are attacked less helps place production services in less attacked ports.

To answer this question, we created the concept of $S$, defined as \textit{the number of services attacked per attacker}. It is independent of the number of honeypots attacked ($H$) and can only take the values $1, 2, 3, 4, 5$ and $S>5$. The limit of $5$ was defined since preliminary results showed that the average number of services attacked per attacker was $5.41$. This question refers to services of the TCP and UDP protocols since other protocols do not have services.

\textbf{The number of attackers per $S$} was calculated as the sum of attackers that attacked $S$ services. The number of \textbf{attacks per $S$} was calculated as the sum of the attacks per attacker that attacked $S$ services. Additionally, we calculated the average, standard deviation and median of the number of attacks per attacker per $S$.

\subsection{How attackers attack?}
\label{sec-how-attackers-attack}

The motivation to understand how attackers attack is to identify patterns that help better evaluate honeypots as early warning systems. Understanding how attackers select servers to attack may help better select the number and location of honeypots to increase the protection of production systems.

The mechanism used to analyse how attackers attack is the sequence of honeypots selected for attack. We want to understand the order of the sequence and the variations. If an attacker sends 1,000 flows to honeypot 1 and then 1,000 flows to honeypot 2, then we consider the order from 1 to 2. We are not interested in the number of flows but only the sequence of honeypots attacked. An attacker that sends 3 flows to honeypot 1, 1 flow to honeypot 2, and 10 flows to honeypot 1 again has the sequence 1,2,1.

For this analysis, we identify each honeypot with a numeric ID (1 to 8) and for each \textbf{attacker} we computed the chronologically sorted sequence of honeypot IDs \textbf{attacked}. Then, we filter out sequences that at least attack all 8 honeypots. An example sequence for a given source IP is $7\rightarrow2\rightarrow6\rightarrow4\rightarrow5\rightarrow8\rightarrow1\rightarrow3\rightarrow6\rightarrow4\rightarrow8\rightarrow1\rightarrow3\rightarrow7$.

As a first step of the sequence analysis, we grouped all sequences that share \textbf{the same starting honeypot ID}. Then we created connected graphs of those sequences. We ended up with eight graphs, one per starting point. Those graphs were used to visually inspect the sequence of honeypots visited by attackers.

In the next step, we used subsequence Dynamic Time Warping (DTW) as a similarity measure to find how often all the possible permutations of length eight occur within all the available sequences and if any visible patterns stand out. DTW~\cite{sakoe1978dynamic} is a similarity measure that is used to compare time series of variable length. The reason to use DTW is that it allows the matching of variable length time series, and it also captures contractions or expansions of time series. For example, a series of honeypot IDs $8 \rightarrow 6 \rightarrow 3$ has zero DTW distance from the sequence $8 \rightarrow 8 \rightarrow 8 \rightarrow 6 \rightarrow 6 \rightarrow 6 \rightarrow 3 \rightarrow 3 \rightarrow3$. In our case, this is a desirable property because an attacker may attack a honeypot more than once before moving to the next one, e.g., in the case of a vertical port scan. Another desirable property of subsequence DTW is that it can also capture cyclic patterns of the form $8 \rightarrow 6 \rightarrow 3 \rightarrow 8 \rightarrow 6 \rightarrow 3 \rightarrow 8 \rightarrow 6 \rightarrow 3$, which occur if attackers repeatedly follow the same pattern of attacks. 

The underlying measure of distance in DTW is the euclidean distance. However, the honeypot IDs should be treated as categorical variables. For our measurements, we assumed that all the distances between nodes are equal to one, i.e., the distance between honeypot 8 and honeypot 7 is the same as between honeypot 8 and honeypot 1. To obtain the distance metrics between sequences, we used the DTW implementation from the library \textit{tslearn}~\cite{tslearn} with the modification of using the constant distance between nodes instead of the Euclidean distance. When we measured how often a subsequence permutation occurs within the attacker sequences, we counted only those that have a distance equal to zero.

\subsection{Profiles of attackers}
\label{sec-profiles-attackers}

One of this paper's most important methodological outputs is the proposal to create attacker profiles based on the preliminary results obtained in the analysis of the number of honeypots attacked ($H$) and the number of services attacked ($S$). 

The number of honeypots attacked $H$, is a discreet number in the range of $1-8$. Given that preliminary results show that most attackers attacked only one honeypot, our profiles differentiated attackers attacking one honeypot or more than one honeypot.

The number of attacked services, $S$, is a discreet number in the range of $1, 2, 3, 4, 5$ and $>5$. Given that preliminary results show that most attackers attack one service, we simplified the profiles to differentiate between attackers that attacked 1 service or more than 1 service.

Our separation gave rise to four high-level types of attacker profiles:
\begin{itemize}
    \item \textbf{Casual Focused}: attacks only one honeypot and only one service ($H=1$, $S=1$).
    \item \textbf{Casual Explorer}: attacks only one honeypot and more than one service ($H=1$, $S>1$).
    \item \textbf{Driven Focused}: attacks more than one honeypot and only one service ($H>1$, $S=1$).
    \item \textbf{Driven Explorer}: attacks more than one honeypot and more than one service ($H>1$, $S>1$).
\end{itemize}

\subsection{Benign scanners}
\label{sec-benign-scanners}

By definition, all connections to a honeypot are considered attacks. However, it is widely known that companies, governments, individuals, and educational institutions actively and repeatedly scan the Internet to search for known vulnerabilities or conduct studies~\cite{scanning:whoiswho,scanning:ncsc,scanning:stanford}. These are known as  \textbf{benign scanners}. To ensure these entities achieve accurate results, they often do not disclose which IPs are used to perform the scans. Therefore, it is common that intelligence collected from honeypots contain \textit{benign} IPs that belong to this group.

To measure the number of benign scanners, we collected 8,508 IP addresses associated with known scanners. These IP addresses were gathered through open-source intelligence tools (OSINT) and third-party platforms\cite{greynoise} during 2023. The collected IPs were compared to the list of attackers in the dataset.

\section{Results}
\label{sec-results}
This section presents the results of applying the research methodology defined in Section~\ref{sec-methodology} to the CTU Hornet 40 dataset. Considering that only 0.076\% of the attackers in the dataset were identified as proxy attackers attacking through the Tor network, we did not remove them from the data. This small percentage of attackers that use Tor as a proxy was also reported in~\cite{kelly2021comparative}.

\subsection{Where do attacks originate from?}
The \textbf{number of attackers} by source continent are shown in~\autoref{tab:continent_attackers_vol}. It can be seen that 54.87\% of the total attackers originate from Asia, 16.84\% from Europe, and 15.01\% from North America. There were significantly fewer attacks from the rest of the continents, with 8.14\% coming from South America, 4.58\% from Africa, and 0.47\% from Oceania. A 0.08\% of attackers did not have a continent associated with their source IPs at the moment of this study. No attackers were found originating from Antarctica (ZZ represents no country defined).

Taking into account the percentage of global Internet penetration rate, 58\% of the attackers are expected to come from Asia, and our results show 54.87\% originate in Asia, meaning we observed 0.94x than expected. 11.2\% of the attackers are expected to come from Africa, and our results show 0.47\%, meaning we observed 0.04x than expected. 13.9\% of the attackers are expected to come from Europe, and our results show 16.84\%, meaning we observed 1.21x than expected. 9.9\% of the attackers are expected to come from South America, and our results show 4.58\%, meaning we observed 0.46x than expected. 6.5\% of the attackers are expected to come from North America, and our results show 15.01\%, meaning we observed 2.3x than expected. 0.6\% of the attackers are expected to come from Oceania, and our results show 0.47\%, meaning we observed 0.78x than expected.


Table~\ref{tab:continent_attackers_vol} also shows the percentage of \textbf{attackers} per continent disaggregated by $H$ (number of honeypots attacked). It can be seen that for all continents, the large majority of attackers attacked only one honeypot. Even though the percentage decreases with the number of honeypots attacked, it actually increases for 8 honeypots.

\begin{table}[tb]
\tiny
\renewcommand{\arraystretch}{1.5}
\caption{Percentage of \textbf{attackers} per continent disaggregated by the number of honeypots attacked}
\begin{center}
\begin{tabular}{c r r r r r r r r r}
\toprule
\makecell[c]{H} & \makecell[c]{\textbf{AF}} & \makecell[c]{\textbf{AS}} & \makecell[c]{\textbf{EU}} & \makecell[c]{\textbf{NA}} & \makecell[c]{\textbf{OC}} & \makecell[c]{\textbf{SA}} & \makecell[c]{\textbf{AN}} & \makecell[c]{\textbf{ZZ}} \\
\midrule
1 &	4.05\% & 	43.30\% & 	11.75\% & 	9.90\% & 	0.40\% & 	6.31\% & 	0\% & 	0.07\% \\ 
2 &	0.28\% & 	5.34\% & 	2.04\% & 	1.40\% & 	0.03\% & 	1.01\% & 	0\% & 	$<$0.01\% \\
3 &	0.13\% & 	2.29\% & 	0.93\% & 	0.64\% & 	0.02\% & 	0.43\% & 	0\% & 	$<$0.01\% \\
4 &	0.06\% & 	1.11\% & 	0.46\% & 	0.37\% & 	0.01\% & 	0.20\% & 	0\% & 	$<$0.01\% \\
5 &	0.03\% & 	0.70\% & 	0.27\% & 	0.28\% & 	0\%    & 	0.10\% & 	0\% & 	$<$0.01\% \\
6 &	0.01\% & 	0.48\% & 	0.24\% & 	0.26\% & 	0\%    & 	0.04\% & 	0\% & 	0\% \\
7 &	0.01\% & 	0.46\% & 	0.17\% & 	0.35\% & 	0\%    & 	0.03\% & 	0\% & 	0\% \\
8 &	0.01\% & 	1.20\% & 	0.98\% & 	1.81\% & 	0.01\% & 	0.03\% & 	0\% & 	0\% \\
\midrule
Total & 4.58\% & 54.87\% & 16.84\% & 15.01\% & 0.47\% & 8.14\% & 0\% & 0.08\% \\
\bottomrule
\end{tabular}
\label{tab:continent_attackers_vol}
\end{center}
\end{table}

The \textbf{number of attacks} per source continent are shown in Table~\ref{tab:continent_attacks_vol}. It can be seen that 41.74\% of total attacks originate from Asia, 36.12\% from Europe, and 19.58\% from North America. The number of attacks in the remaining continents was almost negligible, with 1.48\% of attacks originating from South America, 0.86\% from Africa, and 0.21\% from Oceania. A 0.01\% of attacks originated from attackers that did not have a continent associated with their source IPs at the moment of this study. No attacks were found originating from Antarctica.

Taking into account the percentage of global Internet penetration rate (see~\autoref{sec-where-attacks-from}), we observed 0.72x attacks originating in Asia than expected, 0.08x attacks originating in Africa than expected, 2.6x attacks originating in Europe than expected, 0.15x attacks originating in South America than expected, 3.01x attacks originating in North America than expected, and 0.35x attacks originating in Oceania than expected.

\begin{table}[tb]
\tiny
\renewcommand{\arraystretch}{1.5}
\caption{Percentage of \textbf{attacks} per continent disaggregated by the number of attacked honeypots}
\begin{center}
\begin{tabular}{c r r r r r r r r r}
\toprule
\makecell[c]{H} & \makecell[c]{\textbf{AF}} & \makecell[c]{\textbf{AS}} & \makecell[c]{\textbf{EU}} & \makecell[c]{\textbf{NA}} & \makecell[c]{\textbf{OC}} & \makecell[c]{\textbf{SA}} & \makecell[c]{\textbf{AN}} & \makecell[c]{\textbf{ZZ}} \\
\midrule
1 & 0.56\% & 	32.60\% & 	10.20\% & 	6.05\% & 	0.13\% & 	0.63\% & 	0\% & 	$<$0.01\% \\
2 & 0.07\% & 	1.45\% & 	1.91\% & 	1.44\% & 	0.01\% & 	0.22\% & 	0\% & 	$<$0.01\% \\
3 & 0.06\% & 	0.89\% & 	3.47\% & 	0.44\% & 	0.01\% & 	0.16\% & 	0\% & 	$<$0.01\% \\
4 & 0.03\% & 	0.54\% & 	0.37\% & 	0.42\% & 	0\%    & 	0.12\% & 	0\% & 	$<$0.01\% \\
5 & 0.02\% & 	0.49\% & 	0.38\% & 	0.39\% & 	0.01\% & 	0.10\% & 	0\% & 	$<$0.01\% \\ 
6 & 0.02\% & 	0.43\% & 	8.19\% & 	0.81\% & 	0\%    & 	0.06\% & 	0\% & 	0\% \\
7 & 0.02\% & 	0.75\% & 	2.60\% & 	0.93\% & 	0\%    & 	0.07\% & 	0\% & 	0\% \\
8 & 0.08\% & 	4.59\% & 	9.00\% & 	9.10\% & 	0.04\% & 	0.12\% & 	0\% & 	0\% \\
\midrule
Total & 0.86\% & 	41.74\% & 	36.12\% & 	19.58\% & 	0.21\% & 	1.48\% & 	0\% & 	0.01\% \\
\bottomrule
\end{tabular}
\label{tab:continent_attacks_vol}
\end{center}
\end{table}

Table~\ref{tab:continent_attacks_vol} shows the origin of attacks disaggregated by $H$ (number of honeypots attacked). For all the continents, the majority of attacks go to only one honeypot.

\subsection{How much do attackers attack?}

The distribution of \textbf{attackers} and \textbf{attacks} disaggregated by $H$ (number of honeypots attacked) is shown in~\autoref{tab:howmuchattackersattack}, along with the average of attackers per attackers and its respective standard deviation and median.

\begin{table}[tb]
\tiny
\renewcommand{\arraystretch}{1.5}
\caption{Number of attacks and attackers per number of honeypots attacked ($H$). The percentage is calculated over the respective total amount of attackers and attacks in the dataset}
\begin{center}
\begin{tabular}{l | rr | rr | rrr}
\toprule
\multirow{2}{*}{H} & \multicolumn{2}{c}{\textbf{Number of Attackers}} & \multicolumn{2}{c}{\textbf{Number of Attacks}} & \multicolumn{3}{c}{\textbf{Attacks per Attacker}} \\
 & \multicolumn{1}{c}{Total} & \multicolumn{1}{c}{\%}  & \multicolumn{1}{c}{Total} & \multicolumn{1}{c}{\%}  &  \multicolumn{1}{c}{Avg}  & \multicolumn{1}{c}{Std}  & \multicolumn{1}{c}{Median}  \\
\midrule
1 & 201,183 & 75.80\%  & 2,570,588  & 54.15\%   &  12.78  & 1051.66 &  1\\
2 & 26,780  & 10.09\%  & 222,753    & 4.69\%    &   8.32  & 185.76  &  2\\
3 & 11,775  & 4.44\%   & 218,604    & 4.61\%    &  18.57  & 228.58  &  4\\
4 & 5,834   & 2.20\%   & 64,582     & 1.36\%    &  11.07  & 47.60   &  6\\
5 & 3,674   & 1.38\%   & 60,324     & 1.27\%    &  16.42  & 66.87   &  8\\
6 & 2,763   & 1.04\%   & 421,324    & 8.88\%    & 152.49  & 807.74  & 11\\
7 & 2,710   & 1.02\%   & 190,070    & 4.00\%    &  70.14  & 864.41  & 16\\
8 & 10,704  & 4.03\%   & 998,836    & 21.04\%   &  93.31  & 434.86  & 22\\
 \bottomrule
\end{tabular}
\label{tab:howmuchattackersattack}
\end{center}
\end{table}

Our measurements show that 75.80\% of attackers in the dataset attacked only one honeypot, 10.09\% of attackers attacked two honeypots, 4.44\% of attackers attacked three honeypots, and 4.03\% of attackers attacked all eight honeypots. Overall, 24.20\% of attackers attacked more than one honeypot.

In terms of the number of attacks, 54.15\% of the attacks in the dataset were generated by attackers attacking only one honeypot, 21.04\% was generated by attackers attacking all eight honeypots, and 8.88\% was generated by attackers attacking six honeypots. Overall, 45.85\% of the attacks were generated by attackers attacking more than one honeypot.

As shown in~\autoref{tab:howmuchattackersattack}, the largest variability in the average of attacks per attacker is observed in attackers that attack only one honeypot, with 12.78$\pm$1051.66 attacks per attacker and a median of 1. Attackers that attack six honeypots are those with the highest number of attacks per attacker, with 152.49$\pm$807.74 attacks per attacker and a median of 11.

\subsection{What do attackers attack?}
The average number of destination ports attacked per attacker is 5.44. The distribution of \textbf{attackers} and \textbf{attacks} disaggregated by $H$ (number of honeypots attacked) and $S$ (number of services attacked) is presented in~\autoref{tab:whatattackersattack-volumeattackers} and~\autoref{tab:whatattackersattack-volumeattacks}, respectively.  

\begin{table}[tb]
\tiny
\renewcommand{\arraystretch}{1.5}
\caption{Percentage of attackers per number of attacked services disaggregated by the number of attacked honeypots. The percentage is calculated over the total amount of attackers}
\begin{center}
\begin{tabular}{c |r r r r r r}
\toprule
\makecell[c]{H} & \makecell[c]{\textbf{S$=$1 }} & \makecell[c]{\textbf{S$=$2 }} & \makecell[c]{\textbf{S$=$3 }}  & \makecell[c]{\textbf{S$=$4}}  & \makecell[c]{\textbf{S$=$5 }}  & \makecell[c]{\textbf{S$>$5 }}  \\
\midrule
1 & 73.26\% & 	2.29\% & 	0.12\% & 	0.08\% & 	0.03\% & 	0.19\% \\ 
2 & 8.00\% & 	1.62\% & 	0.22\% & 	0.06\% & 	0.03\% & 	0.13\% \\
3 & 3.21\% & 	0.56\% & 	0.33\% & 	0.16\% & 	0.07\% & 	0.11\% \\
4 & 1.32\% & 	0.31\% & 	0.10\% & 	0.15\% & 	0.12\% & 	0.18\% \\
5 & 0.69\% & 	0.19\% & 	0.06\% & 	0.05\% & 	0.07\% & 	0.34\% \\
6 & 0.40\% & 	0.09\% & 	0.04\% & 	0.03\% & 	0.03\% & 	0.45\% \\
7 & 0.32\% & 	0.06\% & 	0.04\% & 	0.02\% & 	0.02\% & 	0.53\% \\
8 & 1.79\% & 	0.22\% & 	0.14\% & 	0.06\% & 	0.08\% & 	1.66\% \\
\midrule
Total & 89.00\% & 	5.31\% & 	1.05\% & 	0.59\% & 	0.45\% & 	3.59\% \\ 
\bottomrule
\end{tabular}
\label{tab:whatattackersattack-volumeattackers}
\end{center}
\end{table}

\begin{table}[tb]
\tiny
\renewcommand{\arraystretch}{1.5}
\caption{Percentage of attacks per number of attacked services disaggregated by the number of attacked honeypots. The percentage is calculated over the total amount of attacks}
\begin{center}
\begin{tabular}{c |r r r r r r}
\toprule
\makecell[c]{H} & \makecell[c]{\textbf{S$=$1 }} & \makecell[c]{\textbf{S$=$2 }} & \makecell[c]{\textbf{S$=$3 }}  & \makecell[c]{\textbf{S$=$4}}  & \makecell[c]{\textbf{S$=$5 }}  & \makecell[c]{\textbf{S$>$5 }}  \\
\midrule
1 & 31.30\% & 	0.64\% & 	0.04\% & 	0.02\% & 	0.01\% & 	23.60\% \\ 
2 & 1.54\% & 	0.50\% & 	0.10\% & 	0.02\% & 	0.02\% & 	2.33\% \\
3 & 1.06\% & 	0.25\% & 	0.17\% & 	0.09\% & 	0.05\% & 	3.08\% \\
4 & 0.65\% & 	0.13\% & 	0.05\% & 	0.08\% & 	0.08\% & 	0.34\% \\
5 & 0.49\% & 	0.10\% & 	0.06\% & 	0.08\% & 	0.04\% & 	0.50\% \\
6 & 0.46\% & 	0.06\% & 	0.06\% & 	0.03\% & 	0.02\% & 	8.39\% \\
7 & 0.71\% & 	0.09\% & 	0.11\% & 	0.02\% & 	0.02\% & 	2.92\% \\
8 & 2.10\% & 	0.61\% & 	0.37\% & 	0.36\% & 	0.40\% & 	15.88\% \\ 
\midrule
Total	& 38.30\% & 	2.37\% & 	0.96\% & 	0.69\% & 	0.64\% & 	57.04\% \\
\bottomrule
\end{tabular}
\label{tab:whatattackersattack-volumeattacks}
\end{center}
\end{table}

We found that 89\% of attackers attacked only one destination port, 5.31\% of attackers attacked two destination ports, and 3.59\% of attackers attacked more than five destination ports. Overall, 10.99\% of attackers attacked more than one destination port.

We found that 57.04\% of the attacks were generated by attackers attacking more than five destination ports, 38.30\% of the attacks were generated by attackers attacking only one destination port, and 2.37\% of the attacks were generated by attackers attacking only two destination ports. Overall, 61.46\% of attacks were generated by attackers attacking more than one destination port.

Similarly, the average number of attacks per attacker, disaggregated by $H$ and $S$ is presented in~\autoref{tab:whatattackersattack-averageattacks}. Attackers that attack one honeypot and more than 5 destination ports are those with the highest average number of attacks per attacker, with 2,222.12$\pm$21,115.40 attacks per attacker and a median of 16. The highest median number of attacks per attacker corresponds to the attackers that attacked more than 5 ports and all 8 honeypots.

\begin{table*}[tb]
\tiny
\renewcommand{\arraystretch}{1.5}
\caption{Average attacks per attacker based on the number of attackers that attacked a certain number of honeypots and destination ports}
\begin{center}
\begin{tabular}{c|rrr|rrr|rrr|rrr|rrr|rrr}
\toprule
\multicolumn{1}{l}{H} & \multicolumn{3}{c}{\textbf{S=1}} & \multicolumn{3}{c}{\textbf{S=2}} & \multicolumn{3}{c}{\textbf{S=3}} & \multicolumn{3}{c}{\textbf{S=4}} & \multicolumn{3}{c}{\textbf{S=5}} & \multicolumn{3}{c}{\textbf{S\textgreater{}5}} \\
\multicolumn{1}{l}{} & \textbf{avg} & \multicolumn{1}{c}{\textbf{stdev}} & \multicolumn{1}{c}{\textbf{median}} & \textbf{avg} & \multicolumn{1}{c}{\textbf{stdev}} & \multicolumn{1}{c}{\textbf{median}} & \textbf{avg} & \multicolumn{1}{c}{\textbf{stdev}} & \multicolumn{1}{c}{\textbf{median}} & \textbf{avg} & \multicolumn{1}{c}{\textbf{stdev}} & \multicolumn{1}{c}{\textbf{median}} & \textbf{avg} & \multicolumn{1}{c}{\textbf{stdev}} & \multicolumn{1}{c}{\textbf{median}} & \textbf{avg} & \multicolumn{1}{c}{\textbf{stdev}} & \multicolumn{1}{c}{\textbf{median}} \\
\midrule
1 & 7.50  & 114.76 & 1.00  & 4.90  & 8.76   & 2.00  & 5.73  & 27.71  & 3.00  & 4.39   & 1.66   & 4.00  & 5.56  & 3.45   & 5.00  & 2,222.12 & 21,115.40 & 16.00 \\
2 & 3.38  & 9.71   & 2.00  & 5.44  & 21.60  & 2.00  & 7.90  & 22.66  & 3.00  & 6.87   & 11.96  & 4.00  & 8.96  & 8.55   & 9.00  & 310.14   & 1,603.39  & 27.00 \\
3 & 5.78  & 9.83   & 4.00  & 7.72  & 20.94  & 4.00  & 8.78  & 44.79  & 3.00  & 10.27  & 32.07  & 4.00  & 12.08 & 23.11  & 5.00  & 494.46   & 1,379.88  & 20.00 \\
4 & 8.58  & 10.29  & 6.00  & 7.31  & 6.11   & 6.00  & 9.16  & 15.56  & 5.00  & 8.91   & 34.60  & 4.00  & 11.12 & 29.01  & 5.00  & 33.02    & 157.44    & 7.00  \\
5 & 12.52 & 10.97  & 9.00  & 9.64  & 8.60   & 8.00  & 18.01 & 41.99  & 8.00  & 29.36  & 163.19 & 7.00  & 10.91 & 22.26  & 6.00  & 25.79    & 117.87    & 9.00  \\
6 & 20.17 & 102.79 & 11.00 & 12.85 & 14.77  & 10.00 & 23.75 & 60.07  & 11.00 & 21.20  & 59.84  & 9.00  & 12.50 & 11.24  & 8.00  & 331.11   & 1216.36   & 12.00 \\
7 & 38.28 & 211.26 & 13.00 & 24.97 & 37.66  & 14.00 & 46.67 & 195.22 & 16.00 & 18.18  & 16.73  & 14.00 & 17.46 & 14.12  & 14.00 & 96.38    & 1191.19   & 16.00 \\
8 & 20.59 & 102.52 & 8.00  & 49.35 & 147.89 & 16.00 & 47.07 & 61.40  & 29.00 & 104.76 & 349.05 & 32.00 & 94.01 & 236.97 & 35.00 & 167.88   & 652.31    & 54.00\\
\bottomrule
\end{tabular}
\label{tab:whatattackersattack-averageattacks}
\end{center}
\end{table*}

\subsection{How do attackers attack?}

\begin{figure}[!tbp]
    \centering
    \includegraphics[width=0.43\textwidth]{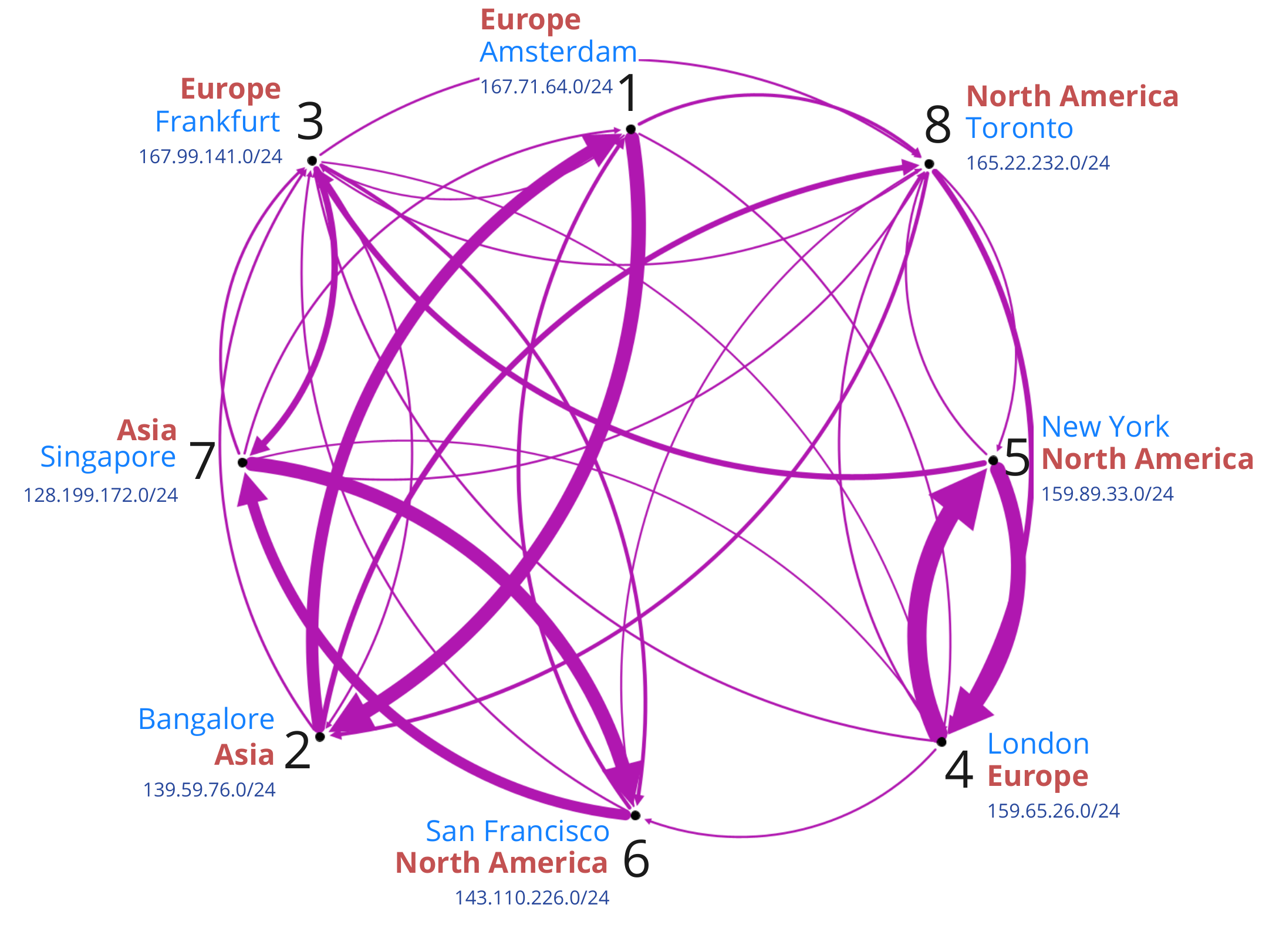}
    \caption{Directed graph of the sequences of attacked honeypots by the subset of attackers that attacked all eight honeypots}
    \label{fig:allnodes}
\end{figure}

The attack patterns were found by analysing the attackers that attacked all 8 honeypots and extracting the sequences in which the honeypots were attacked. These sequences of attacks, built following the steps described in~\autoref{sec-how-attackers-attack}, are represented as a directed connected graph in~\autoref{fig:allnodes}. The edges in the graph are the frequencies of an attacker changing its attention from one honeypot to another. In the figure, it can be seen that certain sequences or loops repeat more than others, such as those between honeypots $4$ and $5$, $1$ and $2$, and $7$ and $6$. This does not mean that the attacker gets into a loop but that, on average, the attackers in one honeypot change focus to the other honeypot with that frequency. Remember that the frequencies are computed for all the sequences that pass through that honeypot.

\begin{figure}[!tbp]
    \centering
    \includegraphics[width=0.43\textwidth]{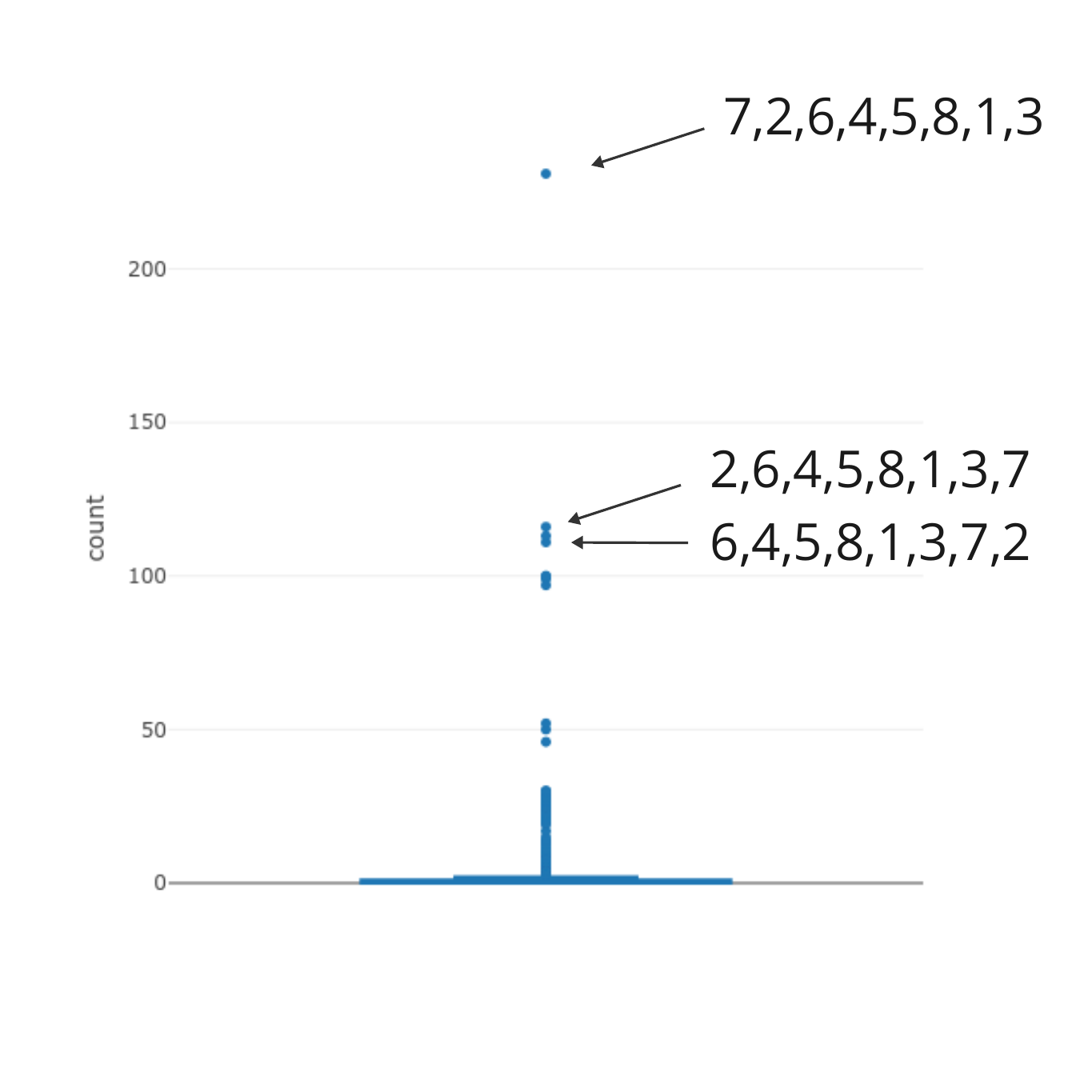}
    \caption{Frequency of attack patterns for each possible subsequence of length eight in the attack paths. The dots represent outlier sequences that appeared more frequently than expected}
    \label{fig:boxplot}
\end{figure}

Using the DTW metric, we counted which was the frequency of each possible subsequence of length eight appearing in the attackers' paths. The results can be seen in~\autoref{fig:boxplot}. The idea is to find if the attacker path includes the same amount of each sequence (a uniform random selection), or if some sequences are \textit{preferred}. Preferred sequences would mean that there is an intentional process of selection. Most subsequences occur only a few times, but a few distinct outliers occur more than 100 times. The eight most frequent occurrences are $7\rightarrow2\rightarrow6\rightarrow4\rightarrow5\rightarrow8\rightarrow1\rightarrow3$ and all its cyclic permutations such as $2\rightarrow6\rightarrow4\rightarrow5\rightarrow8\rightarrow1\rightarrow3\rightarrow7$ and $6\rightarrow4\rightarrow5\rightarrow8\rightarrow1\rightarrow3\rightarrow7\rightarrow6$.

In terms of percentage of the total attack patterns, all subsequences related to $7\rightarrow2\rightarrow6\rightarrow4\rightarrow5\rightarrow8\rightarrow1\rightarrow3$, including its reversals account for 2.16\% of the subsequences. These were chosen as exact matches that had DTW distance metric equal to zero. If we accept as matches patterns that have small differences and increase the threshold distance to 1.0, then the number rises to 6\%. This accounts for subsequences that have two nodes reversed at the beginning or end of the sequence, such as $7\rightarrow2\rightarrow6\rightarrow4\rightarrow5\rightarrow8\rightarrow3\rightarrow1$.

In terms of percentage of the total attack patterns, all subsequences related to $7\rightarrow2\rightarrow6\rightarrow4\rightarrow5\rightarrow8\rightarrow1\rightarrow3$, including its reversals account for 2.16\% of the subsequences. These were chosen as exact matches that had DTW distance metric equal to zero. If we accept as matches patterns that have small differences and increase the threshold distance to 1.0, then the number rises to 6\%. This accounts for subsequences that have two nodes reversed at the beginning or end of the sequence, such as $7\rightarrow2\rightarrow6\rightarrow4\rightarrow5\rightarrow8\rightarrow3\rightarrow1$. 
In terms of percentage of the total attack patterns, all subsequences related to $7\rightarrow2\rightarrow6\rightarrow4\rightarrow5\rightarrow8\rightarrow1\rightarrow3$, including its reversals account for 2.16\% of the subsequences. These were chosen as exact matches that had DTW distance metric equal to zero. If we accept as matches patterns that have small differences and increase the threshold distance to 1.0, then the number rises to 6\%. This accounts for subsequences that have two nodes reversed at the beginning or end of the sequence, such as $7\rightarrow2\rightarrow6\rightarrow4\rightarrow5\rightarrow8\rightarrow3\rightarrow1$. 

\subsection{Profiles of attackers}

\begin{figure}[!tbp]
  \centering
  \begin{subfigure}[b]{0.23\textwidth}
    \includegraphics[width=\textwidth]{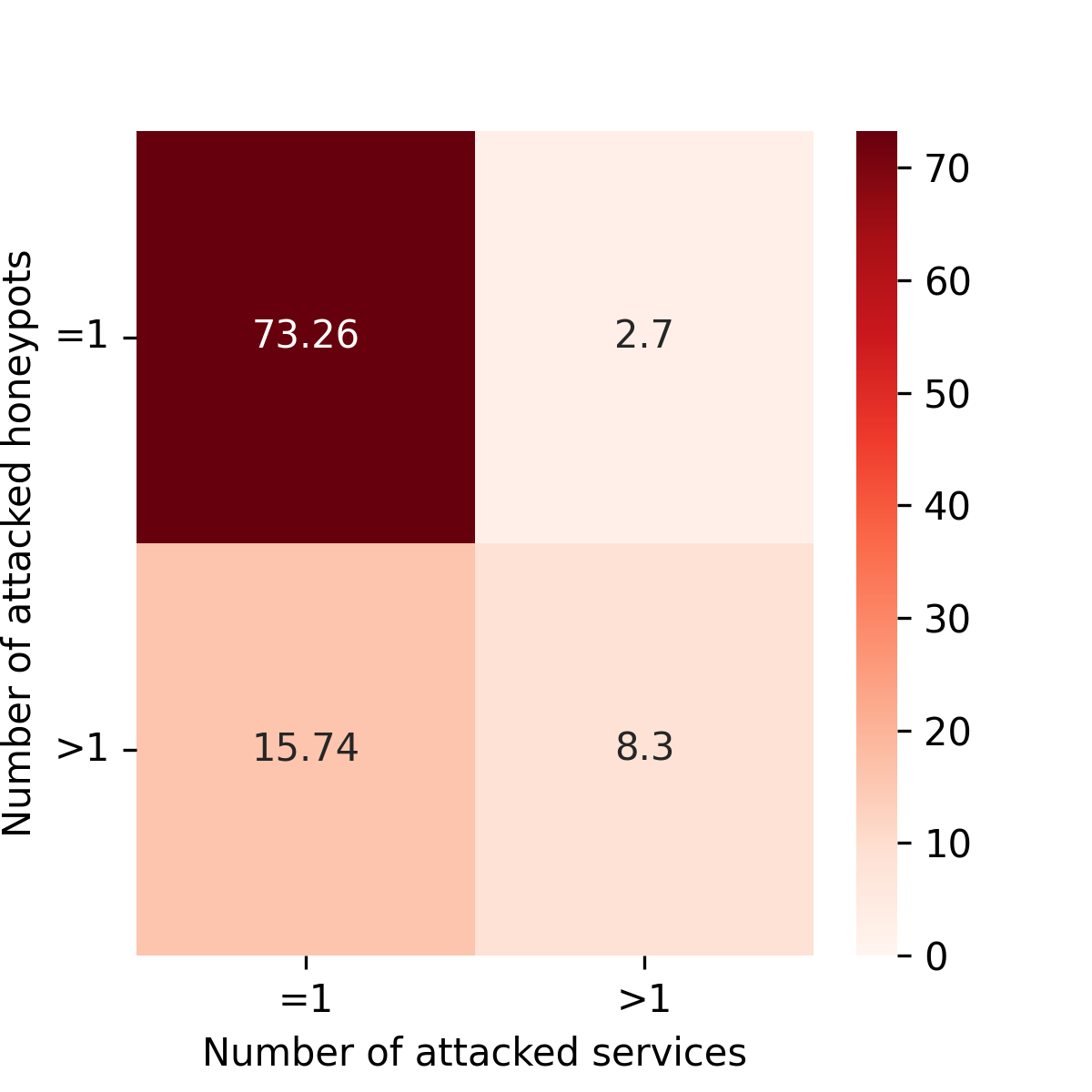}
    \caption{Distribution of attackers}
    \label{fig:matrixattackers}
  \end{subfigure}
  \begin{subfigure}[b]{0.23\textwidth}
    \includegraphics[width=\textwidth]{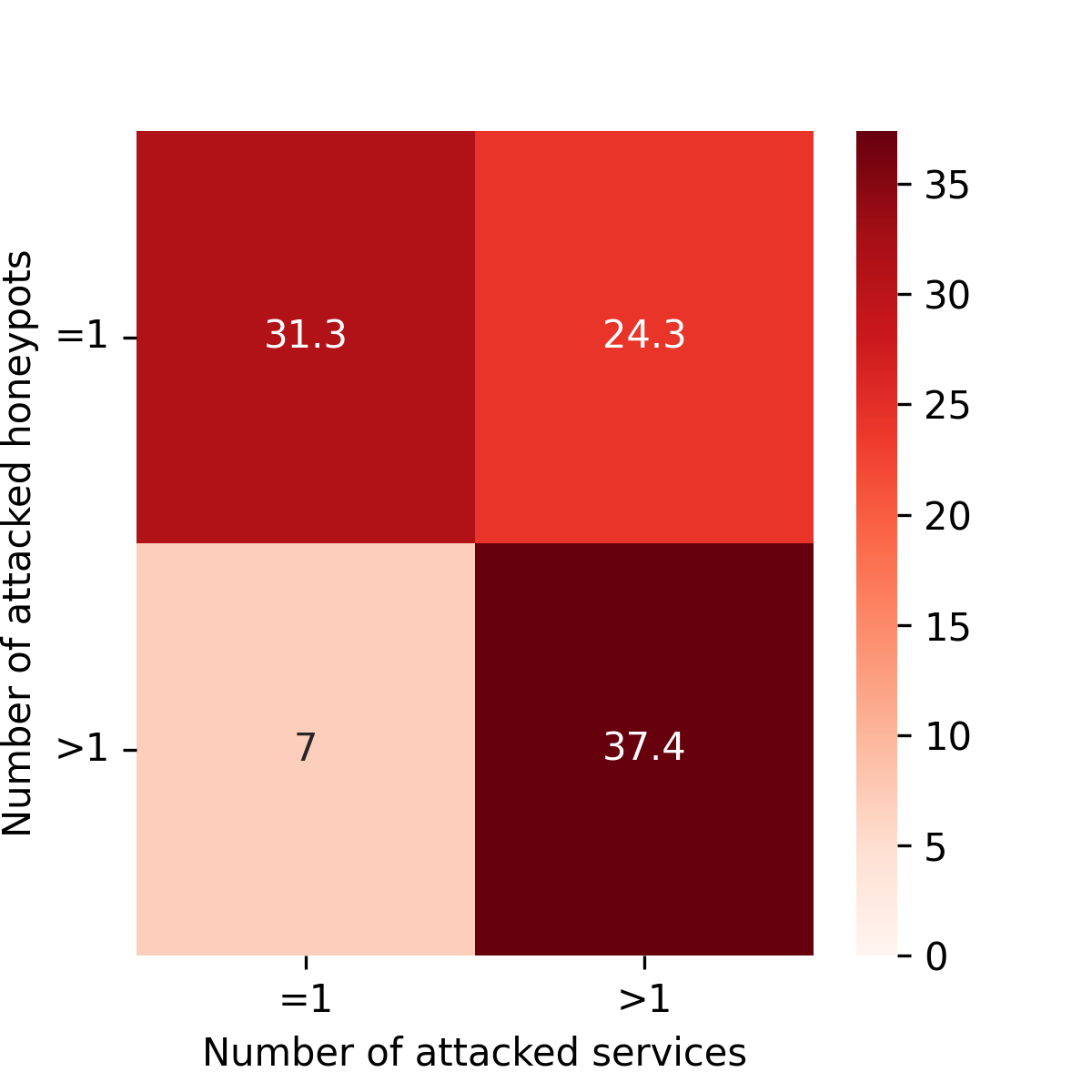}
    \caption{Distribution of attacks}
    \label{fig:matrixattacks}
  \end{subfigure}  
  \caption{Distribution of attackers~(\ref{fig:matrixattackers}) and attacks~(\ref{fig:matrixattacks}) by attacker profile. Casual attackers attack only one honeypot, driven attackers attack more than one. Focused attackers attack only one service. Explorer attackers attack more than one service.}
  \label{fig:matrixprofiles}
\end{figure}

The results of the profiles based on \textit{attackers} and \textit{attacks} are shown in Figure~\ref{fig:matrixprofiles}, separated by 
$H$ (number of attacked honeypots) and $S$ (number of attacked services). This figure shows the four profiles. At the top left ($H=1$, $S=1$) the \textit{Casual focused}, at the top right ($H=1$, $S>1$) the \textit{Causal explorer}, at the bottom left ($H>1$, $S=1$) the \textit{Driven focused}, and at the bottom right ($H>1$, $S>1$) the \textit{Driven explorer}.

According to~\autoref{fig:matrixattackers}, 75.96\% (73.26\% + 2.7\%) of the total attackers can be labelled as \textit{Casual}, attacking only one honeypot. From the total, 24.04\% (15.74\% + 8.3\%) can be labelled as \textit{Driven}, attacking more than one honeypot. From the perspective of the number of destination ports attacked, 89\% (75.26
\% + 15.74\%) of the attackers can be labelled as \textit{Focused}, attacking only one destination port. From the total, 11\% (2.7\% + 8.3\%) of the attackers can be labelled as \textit{Explorer}, attacking more than one destination port.

From the total attackers, 73.26\% can be profiled as \textbf{\textit{Casual focused}}, attacking only one honeypot and one destination port. Casual focused attackers are responsible for 31.3\% of the total attacks in the dataset. Casual focused attackers found to originate from Tor exit nodes were 0.053\%.

From the total attackers, 15.74\% can be profiled as \textbf{\textit{Driven focused}}, attacking more than one honeypot and one destination port. Driven focused attackers are responsible for 7\% of the total attacks. Driven focused attackers found to originate from Tor exit nodes were 0.007\%.

From the total attackers, 2.7\% can be profiled as \textbf{\textit{Casual explorer}}, attacking one honeypot and more than one destination port. Casual explorer attackers are responsible for 24.3\% of the total attacks. Casual explorer attackers found to originate from Tor exit nodes were 0.002\%.

From the total attackers, 8.3\% can be profiled as \textbf{\textit{Driven explorer}}, attacking more than one honeypot and more than one destination port. Driven explorer attackers are responsible for 37.4\% of the total attacks. Driven explorer attackers found to originate from Tor exit nodes were 0.014\%.

\subsection{Benign scanners}

Benign scanners account for 0.63\% of the total number of attackers in the dataset. Broken down by the proposed profiles, 0.42\% of the benign scanner attackers fall in the \textit{driven explorer} profile, 0.17\% in the \textit{casual focused} profile, 0.03\% in the \textit{casual explorer} profile and 0.01\% in the \textit{driven focused} profile. Overall, 3.25\% of the total attacks in the dataset can be associated with benign scanners.

\section{Analysis}
\label{sec-analysis}

Trying to understand the results obtained by processing a large amount of data requires a strong focus on what needs to be answered. This section analyses the original questions and focuses on the results that help answer them.

\subsection{Where do attacks originate from?}

Despite the probably unaccounted seasonal aspects of the dataset, our analysis shows that the majority of the attackers seen in the dataset (in total numbers) originate from Asia. Attackers originating in Asia seem more specialised, with the majority attacking only one honeypot. However, compared with the expected attacks by Internet population, Asia actually sends fewer attacks than expected, and North America sends 2.3x more attacks than expected, with Europe sending 1.21x times more. This is important to put in context the results regarding the number of people connected and not only the raw attacks.

Most of the attackers attack \textbf{one honeypot independently of the continent of origin}. The number of attackers then decreases as the number of attacked honeypots increases. However, there is in general a \textbf{higher number of attackers when attackers attack all eight honeypots}.

Attackers from Africa and South America may appear more efficient in their attacks, as they generate fewer attacks. 4.58\% of attackers originating from Africa generate only 0.86\% of the total attacks. Similarly, 8.14\% of attackers originating in South America generate only 1.48\% of the total attacks. In contrast, attackers from Europe and North America appear more aggressive in their attacks, generating a larger number of attacks per attacker. This is also sustained by the normalised results, in which Europe and North America show more attackers and attacks than expected when taking into account the percentage of global Internet penetration rate.

\subsection{How much do attackers attack?}

Our research shows that honeypots used as early warning systems may be overrated. This is because 75.80\% of attackers attack only one honeypot. This means that if we take one attacker IP address from this group and add it to a blocklist in a production server, the likelihood of it blocking any traffic is very low.

The median of the attackers that connect to only one honeypot ($H=1$) also only sends one flow (one attack). This can be seen in~\autoref{tab:howmuchattackersattack} where the median is 1. The high standard deviation can be explained by a few attackers that scan all 65,535 $TCP$ ports in some honeypots, generating over half a million attacks.

Attackers that attack 6 or more honeypots had the largest median of attacks per attacker (median 11, 16 and 22 for 6, 7 and 8 honeypots). High standard deviations are caused by some attackers scanning more than others.

\subsection{What do attackers attack?}

Attackers show a high degree of specialisation, with 89\% of attackers attacking one service. Which ports attackers attack may be useful for defence independently of how many honeypots are targeted. The high specialisation can be driven by how attackers operate. For example, \texttt{SSH} brute-forcing botnets are in this group, attacking to find servers with weak credentials.

Attackers also show specialisation in the type of application targeted. 5.31\% of attackers target two services. Here we find attackers targeting, for example, the two most common \texttt{TELNET} ports: \texttt{23} and \texttt{2323}, similarly, for \texttt{SSH}, \texttt{FTP} and others.

\subsection{How do attackers attack?}

Attackers that attack all eight honeypots ($H=8$) are of special interest, as they have time and resources to attack all of the Internet and do so repeatedly. Internet-wide scans may take a long time, depending on the tools used and the type of activities done. Our results show two prominent attacking patterns, attackers attacking in ascending order and those attacking in descending order. Many attacker IPs matching these patterns were known in the community for performing brute-forcing attacks. None of these attackers was found to be a benign scanner, and only one was an attacker proxying its attacks through Tor.

\begin{figure}[!tbp]
    \centering
    \includegraphics[width=0.4\textwidth]{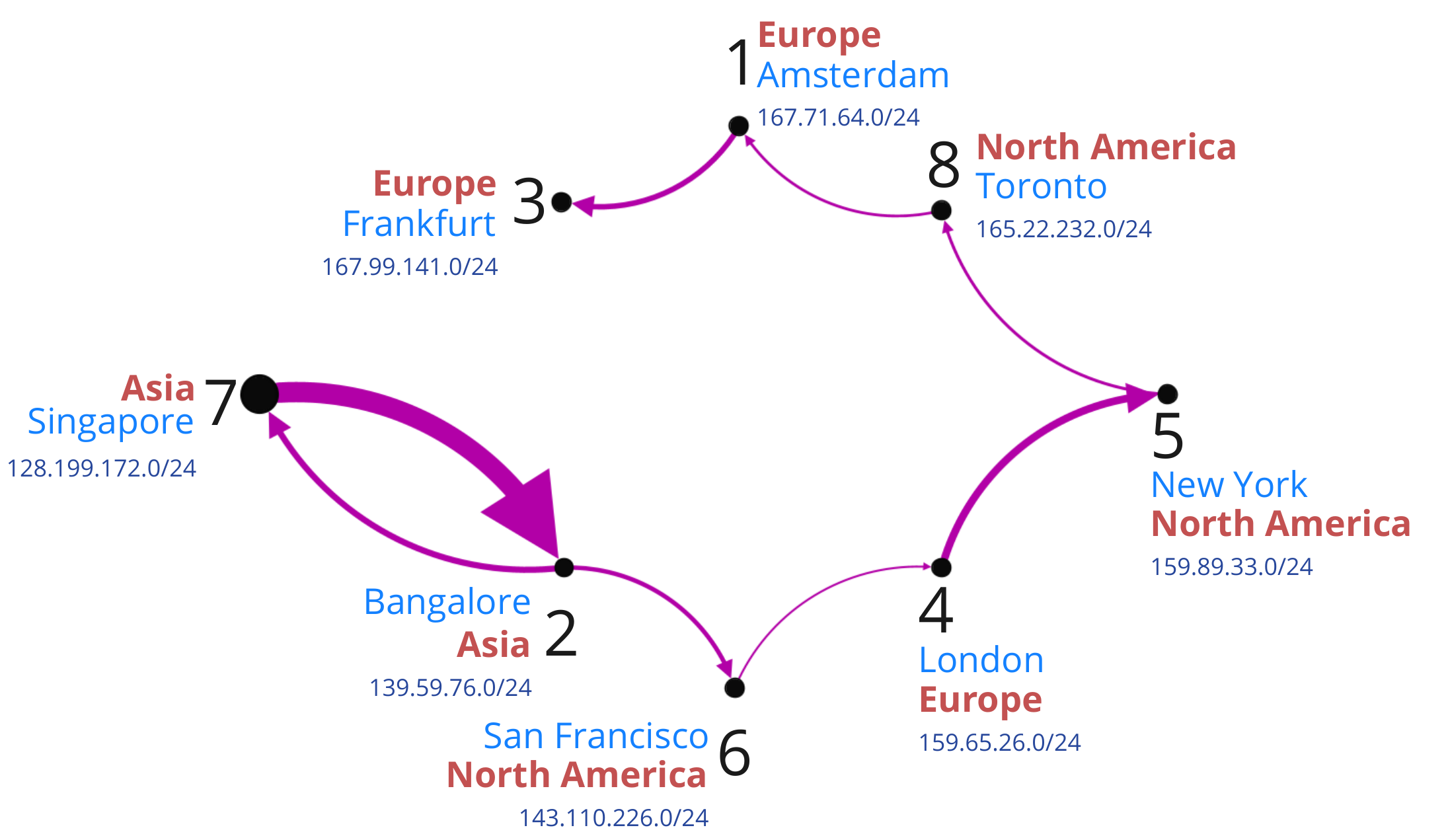}
    \caption{Directed graph illustrating the sequence of honeypots attacked when attackers start by attacking honeypot $7$ and continue to attack all eight honeypots}
    \label{fig:node7start}
\end{figure}

\begin{figure}[!tbp]
    \centering
    \includegraphics[width=0.4\textwidth]{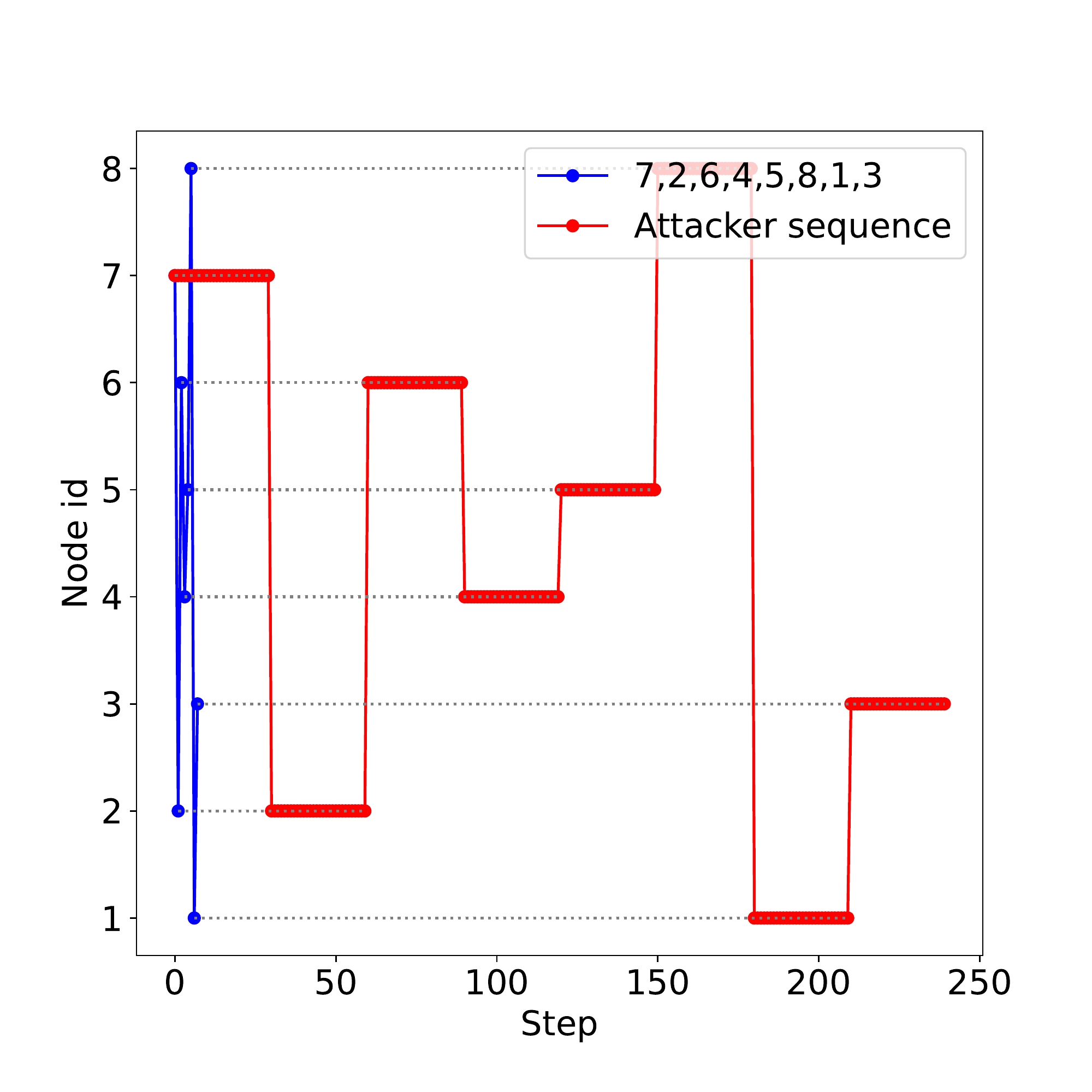}
      \caption{Dynamic Time Warp matching of the pattern 7,2,6,4,5,8,1,3 with one of the attacker sequences that attacked each honeypot 30 times before moving to the next one}
    \label{fig:dtw_pattern}
\end{figure}

One niche of attackers attacks the honeypots sequentially in \textbf{ascending order}. They attack, for example, IP 3.3.3.3, before IP 4.4.4.4.~\autoref{fig:node7start} shows how these attackers start by attacking honeypot 7, which has the lowest IP address number, and move subsequently to other honeypots that have a higher IP address number. The graph shown in~\autoref{fig:node7start} was filtered to only show significant edges in the graph. This pattern was also confirmed using the DTW metric to measure occurrences of subsequences. An example of the type of matches captured by DTW that would not be possible to do without string manipulation is shown in~\autoref{fig:dtw_pattern}.

Another niche of attackers, smaller in size, attacks the honeypots sequentially but in \textbf{descending order}. These attackers start by attacking honeypot 1, which has the highest IP address number, and subsequently move to other honeypots that have a lower IP address number. This pattern was also confirmed using the DTW metric.

Attackers attacking in \textbf{ascending order} are ideal to be used in early warning systems, suggesting that placing a honeypot in the \textbf{lowest possible IP address} and a production server in the \textbf{highest possible IP address} might maximise the available warning time.


\subsection{Profiles of attackers}

We have identified four attacker profiles that we have characterised as \textit{casual} vs \textit{driven} and \textit{focused} vs \textit{explorer}. Casual attackers may become driven attackers over time as more data is being captured. Focused attackers may become explorers for the same reason. However, once an attacker has been \textit{profiled} as driven or explorer, it stays as such for the duration of the study.

Our analysis shows a clear distinction between casual vs driven attackers. Casual attackers are low-scale attackers confined, probably, to specific countries or IP ranges and are not likely to attack globally. Most attackers are casual attackers that attack less than driven attackers. Driven attackers are large-scale attackers with resources to attack globally and repeatedly. Driven attackers attack more on average, but given their reduced numbers, they generate fewer attacks than casual attackers. Driven attackers are of particular interest as they are likely to attack other systems, making them good candidates for blocking.

A clear distinction was also found between focused vs explorer attackers. Focused attackers are very specialised, with a service-oriented interest. In contrast, explorer attackers have a wider and host-centric interest, often attacking groups of services. Explorer attackers are considerably fewer in numbers than focused attackers but are responsible for 61.7\% of attacks.

\subsubsection{Casual focused attackers}

These attackers attack only one honeypot and one destination port. They prefer to attack specific regions or countries instead of globally. They have a service-oriented interest. They may not conduct more widespread attacks due to a lack of resources, time, patience, or interest. Among these may be individuals attempting to discover lost servers, conducting research studies, and certain bots.

According to this study, 73.26\% of the observed attackers can be profiled as \textit{casual focused} attackers. These attackers are not likely to attack another honeypot, \textbf{making blocking these IPs less relevant}. These attackers are not likely to attack another destination port, however, capturing this information may be valuable to evaluate in which ports to place honeypots.

Only 0.17\% of the total attackers were found to be casual focused attackers \textbf{and} benign scanners. Our analysis shows these are often associated with two types of benign scanning entities. One type owns large IP ranges and distributes its scanning activities by assigning a destination IP range and service per source IP. The second type may own smaller IP ranges but performs surgical Internet-wide scans, targeting one service at a time. This second type may turn into driven or explorer attackers over time.

\subsubsection{Casual explorer attackers}

\textbf{\textit{Casual explorer}} attackers attack only one honeypot and more than one destination port. These attackers prefer to attack specific regions or countries instead of globally. They have a wider interest in terms of services targeted, often displaying an interest in the targeted \textit{host} instead of the services themselves. Among these may be people testing tools, certain bots targeting a family of services (e.g., web services may be deployed in ports 80/TCP, 443/TCP and others), and smaller companies focused on a regional threat landscape.

A 2.7\% of the attackers can be profiled as \textit{casual explorer} attackers responsible for 24.3\% of the total attacks observed in the dataset. These attackers are not likely to attack other honeypots, hence using these IPs for blocking would not provide much value. These attackers may attack other services over time, providing good intelligence on where to place services to capture or avoid attacks.

Only 0.03\% of the total attackers were found to be casual explorer \textbf{and} benign scanners. These are often host-discovery scanning entities focused on the most common services attempting to discover new hosts online.

\subsubsection{Driven focused attackers}

\textbf{\textit{Driven focused}} attackers attack more than one honeypot and only one destination port. These attackers attack \textbf{globally} and have a \textbf{service-oriented interest.} These large-scale attackers have resources and time. Among these, we expect to find attackers that own exploits and look for online services to exploit, benign scanning entities attempting to map the global threat landscape, and some narrow-focused IoT botnets and malware. 

Of the total attackers, 15.74\% can be profiled as driven focused attackers. The narrow focus of the attacks generated by driven focused attackers can be seen represented in the volume of attacks associated with them, only 7\% of the total attacks. These attackers are likely to attack other honeypots making their IP addresses a good addition to any blocklist and can be used as an early warning system.

Of the total attackers, 0.01\% were found to be driven focused \textbf{and} benign scanners. We believe these are specialised benign scanners focused on mapping the global status of specific services.

\subsubsection{Driven explorer attackers}

\textbf{\textit{Driven explorer}} attackers attack more than one honeypot and more than one destination port. These attackers attack globally and show a host-oriented interest. These large-scale attackers have resources and time. Among these, we find the bulk of the IoT botnets, wide network attacks, and benign scanners. These attackers are worth observing, collecting intelligence from, and stopping in production.

Of the total attacker, 8.3\% can be profiled as driven explorer attackers. The wide interest of these attackers can be seen in the volume of attacks they generate, as high as 37.4\% of the total attacks. This also means that these attackers may attack more honeypots and services and attack repeatedly. The IP addresses and services associated with these attackers should have high value in preventing attacks on production servers. 

Of the total attackers, 0.42\% were found to be driven explorer \textbf{and} benign scanners. This is how benign scanners are expected to behave. Our analysis shows that under this profile, we find the bulk of IP addresses associated with well-known benign scanners such as ShadowServer, Censys, Shodan, Google and others.

\section{Recommendations}
\label{sec-recommendations}

Defenders interested in using honeypot data on IPs and services as intelligence in early warning systems for protecting production services should focus on \textit{driven focused} and \textit{driven explorer} attackers. With at least two geographically dispersed honeypots, it is possible to identify these attackers and block them in a production system.

Defenders interested in mapping and studying internet attacks \textit{globally} should deploy geographically dispersed honeypots. The number of honeypots deployed will impact the total information collected for statistics. But not all attackers and attacks should be taken to have a \textit{global} impact.

Defenders interested in learning which ports should be avoided for production services should focus on casual explorer and driven explorer attackers. One or more honeypots should allow identifying these attackers.

\section{Limitations and Future Work}
\label{sec-futwork}

One key limitation of this work is the use of a dataset of passive honeypots network traffic. Previous work remarks on the importance of interaction in honeypots, which needs to be considered. We aim to address this in future work by expanding our analysis to other datasets.

Additionally, the dataset comprises only IPv4 network data; future work will attempt to perform similar measurements, including IPv6 addresses. At the moment, attackers do not seem to be using IPv6 extensible, so we believe the measurement of attacks did not lose relevance.

Future work will include a comprehensive verification of proxy servers to quantify the attackers actively trying to hide their origin through reverse DNS and other techniques. Comparing the results with other existing datasets in the area is also part of our future work, to help validate the results obtained in this research.

\section{Conclusions}
\label{sec-conclusions}

This research explored attack patterns in a network dataset of eight geographically dispersed honeypots driven by four core questions: where do attacks originate from and how much, what, and how do attackers attack? The answers to these questions were used to create four attackers' profiles. Ultimately, we explored the results to answer find out how important the role played geolocation in honeypot placement, specifically when it comes to using honeypots as early warning systems.

Through behavioural analysis, we created four attacker profiles that led to a better understanding of the value of the information collected from honeypots and its usability for defence purposes. Our research shows that driven explorer attackers have more attacking power, displaying more attacks per attacker overall. These attackers are worth observing, collecting intelligence from, and stopping in production.

Further study of these behavioural profiles led to identifying two prominent attack patterns for driven explorer attackers, those attacking the Internet sequentially in ascending order and those attacking the Internet sequentially in descending other. 

Analysis indicates that as few as two distributed honeypots can be useful in helping organisations identify the most aggressive and driven attackers and efficiently defend their production servers against them. 

Finally, more research needs to be done to evaluate how honeypot software affects the influx of attacks on honeypots over time. The identification and profiling of benign scanners can also significantly help increase the quality of the data collection from honeypots and deserves further study.

\section*{Acknowledgements}

This work was partially supported by the Strategic Support for the Development of Security Research in the Czech Republic 2019--2025 (IMPAKT 1) program, by the Ministry of the Interior of the Czech Republic under No. VJ02010020 -- AI-Dojo: Multi-agent testbed for the research and testing of AI-driven cyber security technologies.

\bibliographystyle{plain}
\bibliography{references}

\end{document}